\documentclass[twocolumn,superscriptaddress, longbibliography,
amsmath,amssymb,
]{revtex4-2}

\usepackage{graphicx}
\usepackage{dcolumn}
\usepackage{bm}
\usepackage{hyperref}
\usepackage{xcolor}
\usepackage[normalem]{ulem}

\begin{document}

\preprint{APS/123-QED}

\title{Leveling of MHD turbulence imbalance in shear flows}


\author{Mariami Kavtaradze}	
\affiliation{Abastumani Astrophysical Observatory, Abastumani 0301, Georgia}
	\affiliation{Department of Physics, Faculty of Exact and Natural Sciences, Tbilisi State University, Tbilisi 0179, Georgia}
	\author{George Mamatsashvili} \email{g.mamatsashvili@hzdr.de}
    \affiliation{Helmholtz-Zentrum Dresden-Rossendorf, Bautzner Landstr. 400, D-01328 Dresden, Germany}
\affiliation{Abastumani Astrophysical Observatory, Abastumani 0301, Georgia}
	\author{George Chagelishvili}
\affiliation{Abastumani Astrophysical Observatory, Abastumani 0301, Georgia}
    \affiliation{M. Nodia Institute of Geophysics, Tbilisi 0193, Georgia}
    \author{Elene Uchava}
\affiliation{Abastumani Astrophysical Observatory, Abastumani 0301, Georgia}

\date{\today}

\begin{abstract}
We investigate magnetohydrodynamic (MHD) turbulence in plane shear flows with a streamwise background magnetic field in the super-Alfv\'enic regime, i.e., when the flow velocity is larger than the local Alfv\'en speed. In this regime, turbulence is energetically supplied by large-scale shear of the flow via transient, or non-modal amplification of Alfv\'en waves, which are the building blocks of MHD turbulence,  and streamwise-uniform, mostly nonmagnetic modes. We show that shear reduces  turbulence imbalance, driving the system toward a balanced state -- the energies of counter-propagating Alfv\'en waves become essentially equal,  even  for an initially perfectly imbalanced Alfv\'enic turbulence.  This balancing occurs due to the shear-induced linear non-modal dynamics of Alfv\'en waves -- their non-modal growth, resulting in over-reflection, and coupling between counter-propagating wave branches. Because of its linear origin, this process of balancing of MHD turbulence by shear is fundamentally different from the nonlinear processes operating  in standard MHD turbulence in the absence of a mean shear flow.  It can have direct implications for understanding the balanced/imbalanced nature of MHD turbulence in shear regions of the solar wind.
\end{abstract}

\maketitle


\section{Introduction}\label{sec:intro}

Turbulence in magnetized plasmas, often described within magnetohydrodynamics (MHD), is widespread in natural and laboratory settings. Its first theoretical model was proposed by Iroshnikov \cite{Iroshnikov1964} and Kraichnan \cite{Kraichnan1965} as an extension of Kolmogorov’s \cite{kolmogorov1941} theory of isotropic hydrodynamic turbulence, incorporating a magnetic field but still assuming isotropy. Later, Goldreich and Sridhar \cite{Goldreich-Sridhar1995} dropped this limitation and addressed anisotropy of MHD turbulence, which is  naturally caused by a sufficiently strong mean background magnetic field. As is well known in incompressible MHD, such a mean field supports Alfv\'en waves with two distinct polarizations: pseudo-Alfv\'en and shear-Alfv\'en waves.  Each polarization consists of counter-propagating wave components. A key factor in MHD turbulence is the difference between the amplitudes of these counter-propagating Alfv\'en waves, which determines the balanced or imbalanced nature of the turbulence and has been the subject of extensive study over the past decades (e.g., \cite{cho-2000, Maron2001, lithwick-2003, lithwick-2007, perez-2010, beresnyak-2008, beresnyak-2010}). The growing interest in this area has been largely motivated by observations of MHD turbulence in the solar wind (e.g. \cite{belcher-1971, smith2005, zhao-2022, yang-2023}). Studying the degree of imbalance in MHD turbulence is essential to understand energy and momentum transfers across scales (e.g., \cite{lithwick-2007, beresnyak-2008, perez-2012, yokoi-2023}) as well as heating and particle acceleration \cite{teaca-2014}. These processes are crucial for interpreting astrophysical and heliospheric (e.g., solar wind) plasma observations.

In these settings, magnetized plasma,  as a rule,  exhibits large-scale inhomogeneous kinematics, or shear flows that act as a significant source of dynamical activity.  For example, shear flows in the solar wind have long been recognized as a source of turbulence therein \cite{coleman-1968, Roberts-1987a, Roberts-1987b, roberts-1992, Goldstein1999,Matthaeus-2004, Breech_2005, Breech_2008}. Of course,  in astrophysical plasma flows, complex thermodynamic effects are at work alongside kinematic ones. Nevertheless, shear is often the main source of energy for turbulence in the flow.  So, omitting thermodynamic factors for the moment,  we aim to grasp shear-driven processes that constitute the core of turbulence dynamics.

Shear flow -- one of the central drivers of plasma dynamics -- is a likely factor determining the balanced/imbalanced nature of MHD turbulence as well. Indeed, observations of the solar wind indicate that, despite being mostly imbalanced, regions of balanced turbulence also exist that coincide with strong velocity shear regions \cite{Roberts-1987a, Roberts-1987b, Bavassano_1998, Borovsky_2016, Borovsky2019,  lovieno-2015, parashar-2020, Soljento-2023}. This has motivated us to investigate in more detail the physics underlying the effect of shear on the degree of imbalance of MHD turbulence. 

Previous studies of forced, strong, imbalanced MHD turbulence \cite{perez-2010, perez-cross-helicity-2010} showed that numerical simulations are sensitive to the forcing method, complicating the description of the inertial range and energy spectra in Fourier space. However, in shear flow turbulence, the inertial range, which is a range of action of nonlinear processes only, is strictly speaking absent, because both shear-induced linear and nonlinear processes operate in parallel over a broad range of length-scales \cite{horton2010, mamatsashvili2014, mamatsashvili2016}. This naturally calls into question the applicability of the inertial range concept in shear flows. 

Shear-induced linear dynamical processes originate from the non-self-adjointness, or non-normality of the mathematical operators governing linear evolution of perturbations in smooth shear flows without inflection points. The corresponding eigenfunctions of non-normal operators are not mutually orthogonal due to shear and can therefore strongly interfere \cite{Reddy_etal1993, Trefethen_etal1993}. Because of this interference, a particular solution composed of a superposition of eigenfunctions can still exhibit substantial growth over a finite time interval even when all eigenfunctions are modally stable, i.e., contain no exponentially growing components and decay in time   (see e.g.,  \cite{Schmid-Henningson2001, Schmid_2007, Kerswell_2018} for hydrodynamic and  \cite{Fraser_2026} for MHD shear flows as well as  references therein). This phenomenon, caused by flow shear, is referred to as transient, or non-modal growth of perturbations. It occurs over a broad range of length-scales, depends on the orientation of the wavevector of perturbation harmonics  relative to the shearwise direction and hence is anisotropic in Fourier space \cite{chagelishvili2016, mamatsashvili2016, Gogichaishvili-2017, Gogichaishvili-2018}. It is regarded as an important mechanism for triggering transition to turbulence in shear flows that are otherwise modally stable \cite{Grossmann2000, Schmid-Henningson2001, Schmid_2007, Kerswell_2018}. However, due to its transient character, the non-modal growth itself  has a drawback: it requires nonlinear positive feedback to continuously operate and energetically supply turbulence \cite{Gebhardt_Grossmann1994,Baggett_etal1995,Waleffe1995,Hamilton1995,Grossmann2000}.  

The key role of nonlinearity in the sustenance of subcritical turbulence and, in particular, its interplay with linear non-modal dynamics was subsequently studied in detail. Based on low-dimensional models of wall-bounded hydrodynamic shear flows, a nonlinear cycle underlying the self-sustaining process (SSP) was identified and described in \cite{Waleffe1995,Waleffe1997}. This cycle consists of three basic interacting elements: (1) streamwise rolls which, by redistributing momentum, lead to the non-modal growth of (2) streaks, i.e., spanwise modulations of the mean streamwise velocity of the flow, via the lift-up mechanism; these streaks, in turn, become unstable via (3) an inflectional, or secondary instability. The nonlinear self-interaction of this secondary instability provides positive feedback that regenerates the rolls, thereby closing the self-sustaining cycle. This SSP, which is widely accepted in the fluid dynamics community, is therefore based on the dynamical interplay of coherent roll-streak structures in physical space, which are the characteristic features of wall-bounded shear flows, such as plane Couette flow, channel flow and boundary layer flow. In this SSP, the main emphasis is placed on the role of nonlinearity, while boundaries/walls are essential because they support coherent roll–streak structures and determine the near-wall regeneration cycle \cite{Hamilton1995}.

Another type of SSP operating in unbounded or effectively boundary-free modally stable hydrodynamic shear flows (as is usually the case in many natural flows) was proposed in \cite{Gebhardt_Grossmann1994, Baggett_etal1995} using again a simplified reduced order model that does not rely on rigid boundaries. It is based on the subtle interplay of linear non-modal growth of perturbations and nonlinear processes. Specifically, nonlinearity, conserving the total energy, redistributes (mixes) energy among modes and continuously replenishes ``misfit'' perturbations oriented relative to the base flow in such a way that the latter can undergo non-modal growth. Thus, in this SSP of unbounded modally stable shear flows, the linear non-modal dynamics plays a central role because it enables perturbations to extract energy from the base flow and inject it into turbulence.

This SSP, based on the interplay between linear non-modal growth and nonlinear processes, was further studied beyond low-dimensional models using direct numerical simulations of fully developed turbulence in modally stable hydrodynamic shear flows, which account for spectral anisotropy in Fourier space \cite{Grossmann2000,Rempfer2003,Umurhan2004,horton2010,mamatsashvili2016}. In particular, using the model of homogeneous shear turbulence, it was shown in \cite{mamatsashvili2016} that: (i) a generic nonlinear process in shear flows is an anisotropic redistribution of modes over wavevector orientation in Fourier space -- \textit{the nonlinear transverse cascade} -- which is caused by shear and is thus topologically distinct from conventional (inverse/direct) nonlinear cascades and (ii) the transverse cascade provides a positive feedback by continually replenishing non-modally growing modes, closing the sustenance loop. In other words, turbulence is self-sustained through a cooperative and refined interplay between linear non-modal growth and nonlinear transverse cascade. This process was also shown to be responsible for the sustenance of turbulence in MHD shear flows \cite{mamatsashvili2014, Gogichaishvili-2017,  Mamatsashvili2020,held2022}. 

In this paper, we investigate the dynamics of MHD turbulence in plane shear flows subject to a background magnetic field, focusing on its balanced/imbalanced nature. We consider a super-Alfv\'enic regime, that is, when the flow velocity exceeds the Alfv\'en speed associated with the field, so that shear-induced non-modal processes play a dominant role in turbulence sustenance and dynamics. Our main result is that sufficiently strong shear can reduce the degree of turbulence imbalance, equalizing the amplitudes of counter-propagating Alfv\'en waves even when these amplitudes differ substantially initially.  This process of turbulence balancing is governed by the linear non-modal dynamics of Alfv\'en waves, which is based on two basic shear-induced processes: (i) non-modal growth of Alfv\'en waves, resulting in wave over-reflection, and (ii) linear coupling between different Alfv\'en wave branches, whereby an initially excited wave branch generates a counter-propagating one \cite{Chagelishvili1996, Chagelishvili_etal1997,Poedts1998, Gogoberidze2004, Gogoberidze2007,Gogichaishvili2014}.

The essence of the over-reflection phenomenon of Alfv\'en waves in unbounded shear flows is as follows \cite{Gogichaishvili2014}. Consider an incident Alfv\'en wave propagating at a certain angle to the flow, when its wavenumbers $k_y$ in the streamwise ($y$) and $k_x$ in the shearwise ($x$) directions have different signs, $k_x/k_y<0$ (assuming the flow with velocity $U_{0y}$ along the $y$-axis and shear $S=-dU_{0y}/dx>0$ along the $x$-axis). The shear of the flow leads to the non-modal growth of the wave amplitude and makes the shearwise wavenumber change with time, $k_x(t)$,  so that the latter decreases by absolute value, becomes zero and changes sign. At the time when $k_x(t)$ approaches and crosses zero, another branch of the Alfv\'en wave emerges due to the mode coupling process, which propagates opposite to the incident wave. This counter-propagating branch is referred to as the reflected wave, while the incident wave becomes a transmitted wave with $k_x(t)$ and $k_y$ having now the same signs, $k_x(t)/k_y>0$. As a result of non-modal growth, both of these wave branches -- the transmitted and reflected waves -- attain amplitudes larger than that of the incident wave by extracting energy from the base flow, hence this process is referred to as wave over-reflection (see Sec. II A). Thus, in this case of an unbounded shear flow, transmission and reflection of Alfv\'en waves occurs in fact in time: initially there is only an incident wave, which later, when shearwise wavenumber changes sign, is converted into transmitted and reflected waves.

In this paper, we present a novel view on the nature of balance/imbalance of MHD turbulence in shear flows  based on the shear-induced linear processes -- non-modal growth, resulting in the wave over-reflection, and mode coupling -- of Alfv\'en waves discussed above rather than by their nonlinear interactions only as is the case in classical, i.e., shearless and externally forced, MHD turbulence.

The paper is organized as follows: The physical model of the magnetized shear flow, the main equations as well as the linear dynamics of pseudo- and shear-Alfv\'en waves are described in Sec. II.  Numerical setup, general nonlinear evolution,  demonstration of the balancing of MHD turbulence by shear and energy spectra are presented in Sec. III. Conclusions are given in Sec.  IV.

\section{Physical model}

We consider an unbounded isentropic plane flow along the $y$-axis with a constant shear $S>0$ of velocity along the $x$-axis, ${\bm U}_0=-Sx\bm{e}_y$, in the Cartesian coordinate system $(x,y,z)$, where $\bm{e}_y$ is the unit vector along the $y$-axis. A uniform background magnetic field ${\bm B}_0=B_{0y}\bm{e}_y,~B_{0y}>0$ is applied parallel to the flow. Owing to its kinematic and thermodynamic simplicity, this basic model of a magnetized flow with constant shear provides a useful framework for isolating and understanding the effects of large-scale velocity shear on MHD turbulence, with relevance, for example, to shear regions in the solar wind \cite{Poedts1998, Gogoberidze2007, Hollweg2013}. The perturbations of velocity ${\bm u}$ and magnetic  field ${\bm b}$ about this base flow are governed by the equations of incompressible non-ideal MHD  
\begin{multline}\label{eq:mom}
\frac{D{\bm{u}}}{D t}+({\bm{u}}\cdot\nabla){\bm{u}}-Su_x\bm{e}_y=-\frac{1}{\rho}\nabla P\\+(\bm{U}_A\cdot\nabla)\bm{b}+(\bm{b}\cdot\nabla)\bm{b}+\nu\Delta\bm{u},
\end{multline}
\begin{equation}\label{eq:ind}
    \frac{D\bm{b}}{D t}=-Sb_x\bm{e}_y+(\bm{U}_A\cdot\nabla)\bm{u}+\nabla\times(\bm{u}\times\bm{b})+\eta\Delta\bm{b},
\end{equation}
\begin{equation}\label{eq:divu}
\nabla\cdot\bm{u}=\nabla\cdot\bm{b}=0,
\end{equation}
where $\rho$ is the constant density, $P$ is the total pressure, equal to the sum of the thermal and magnetic pressures, and $\bm{U}_A=\bm{B}_0/(4\pi\rho)^{1/2}$ is the Alfv\'en speed corresponding to the background field.  $\nu$ is the kinematic viscosity and $\eta$ is the Ohmic resistivity.  Here $D/Dt\equiv \partial/\partial t-Sx\partial/\partial y$ is the advective derivative along the base flow. The magnetic field perturbation is measured in velocity units, $\bm{b}/(4\pi\rho)^{1/2}\rightarrow \bm{b}$. Following a standard approach in MHD turbulence studies \cite{Goldreich-Sridhar1995, Maron2001, perez-2010}, we rewrite Eqs. (\ref{eq:mom})--(\ref{eq:divu}) in terms of Els\"asser variables, $\bm Z^{\pm} = \bm{u} \pm\bm{b}$:
\begin{multline}\label{eq:Elsassers}
\left(\frac{D}{D t} \mp \bm{U}_A\cdot\nabla\right)\bm{Z}^{\pm} +(\bm{Z}^{\mp}\cdot\nabla)\bm{Z}^{\pm} =-\frac{1}{\rho}\nabla P + SZ_x^{\mp}\bm{e}_y\\+\frac{\nu+\eta}{2}\Delta\bm{Z}^{\pm}+\frac{\nu-\eta}{2}\Delta\bm{Z}^{\mp},
\end{multline}
\begin{equation}\label{eq:divZ}
\nabla \cdot {\bm Z}^{\pm}=0.
\end{equation}
The main novelty in Eq. (\ref{eq:Elsassers}) compared to the standard equations for ${\bm Z}^{\pm}$ used in MHD turbulence theory in the shearless case ($S=0$) is the term $SZ_x^{\mp}\mathbf{e}_y$ arising from shear $S$ on the right hand side. This term plays a key role in the present case with shear -- it is responsible for the linear coupling of counter-propagating Alfv\'en waves, characterized by ${\bm Z}^{\pm}$, as well as for their energy exchange with the base flow resulting in the non-modal growth, which we describe in more detail in Sec. II A below.  By contrast, in the classical case without a base shear flow, these waves are coupled through the nonlinear term $(\bm{Z}^{\mp}\cdot\nabla)\bm{Z}^{\pm}$ on the left hand side of Eq.  (\ref{eq:Elsassers}) and through the part of the diffusion term on the right hand side that is proportional to $\nu-\eta$. From these two coupling terms, the nonlinear one is usually considered dynamically far more important than the dissipation one, since it determines nonlinear cascades and energy spectra of MHD turbulence \cite{Maron2001, Boldyrev_2006PhRvL}. Below we take for simplicity $\nu=\eta$, when this diffusion coupling term vanishes.  Our goal is to analyze the shear-induced linear non-modal dynamics of Alfv\'en waves and its effect on the imbalance/balance properties of MHD turbulence in shear flows. 

We stress that shearless MHD turbulence considered in most studies is fueled by some external forcing (e.g.,\cite{Goldreich-Sridhar1995, boldyrev-2009, perez-2010, beresnyak-2008}), whereas here we study the turbulence with an intrinsic energy source -- the base MHD shear flow ${\bm U}_0$, as is usually the case in many natural settings. Although this constant shear flow subject to a uniform streamwise field is modally stable,  i.e., it has no exponentially growing modes  according to the modal analysis \cite{Romanov-1973, Stern-1963, Ogilvie-1996}, perturbations (Alfv\'en waves) can still undergo non-modal growth due to the term $SZ_x^{\mp}\bm{e}_y$ proportional to shear,  which,  as noted above, provides their main supply of energy, extracting it from the base flow \cite{Gogoberidze2004,mamatsashvili2014, Gogichaishvili2014}.

One of the central characteristics of MHD turbulence is the total cross-helicity 
\[
H_c=\int (\bm{u}\cdot \bm{b})dV=\frac{1}{4}\int \left[(Z^{+})^2-(Z^{-})^2\right]dV,
\]
where the integral is taken over the entire flow domain, which measures the correlation of the velocity and magnetic field perturbations, often referred to as Alfv\'enicity in solar wind studies \cite{Matthaeus-2004}. Equivalently, it characterizes the degree of turbulence imbalance, that is, the difference between the Els\"asser energies $(1/4)\int (Z^{+})^2 dV$ and $(1/4)\int (Z^{-})^2 dV$ carried by Alfv\'en waves propagating in the opposite directions. In the strongly imbalanced case, one of the Els\"asser energies is much larger than the other. In the ideal case without shear, $H_c$ is conserved if perturbations obey periodic boundary conditions or vanish at infinity \cite{moffatt1978}, implying the conservation of the imbalance/balance of turbulence.  However, in shear flows $S \neq 0$, as follows from Eqs.  (\ref{eq:mom})-(\ref{eq:divu}), $H_c$ is no longer conserved and varies in time at a rate proportional to shear
\[
\frac{dH_c}{dt}=S\int(u_xb_y-u_yb_x)dV=\frac{S}{2}\int(Z_x^{-}Z_y^{+}-Z_x^{+}Z_y^{-})dV,
\]
which can have consequences for the turbulence imbalance, as it will be shown below.

Following \cite{perez-cross-helicity-2010}, we also introduce the normalized cross-helicity $\sigma_c=H_c/E$, where $E=(1/2)\int (u^2+b^2)dV=(1/4)\int[ (Z^{+})^2+(Z^{-})^2]dV$ is the total energy of perturbations in the domain equal to the sum of the kinetic, $(1/2)\int u^2 dV$,  and magnetic, $(1/2)\int b^2 dV$, energies.  It is often used as a convenient diagnostic measure of imbalance, or Alfv\'enicity in the solar wind MHD turbulence \cite{Bavassano_1998,malara-2001,Matthaeus-2004}. Defining a volume-average over the domain, $\langle \cdot \rangle=(1/V)\int \cdot~dV$, where $V=L_xL_yL_z$ is the domain volume, the normalized cross-helicity can be written as
\begin{equation}\label{eq:sigma}
\sigma_c=\frac{2\langle \bm{u}\cdot \bm{b}\rangle}{\langle u^2+b^2\rangle}=\frac{\left\langle( Z^{+})^{2}\rangle -\langle(Z^{-})^{2}\right\rangle}{\left\langle (Z^{+})^{2}\rangle +\langle(Z^{-})^{2}\right\rangle}.
\end{equation}
When the velocity and magnetic field perturbations exhibit a strong correlation (alignment) of a preferred sign, $\bm{b}\approx \pm \bm{u}$, i.e., high Alfv\'enicity, the turbulence is strongly imbalanced and $\sigma_c$ is close to unity by absolute value, $\sigma_c\approx \pm 1$. By contrast, in the balanced case, counter-propagating Alfv\'en waves carry equal amounts of energy, the velocity and field perturbations are not correlated and hence $\sigma_c\approx 0$.

\subsection{Linear dynamics and over-reflection of Alfv\'en waves}

To classify perturbation modes present in the flow,  we consider first the linear regime and decompose the  quantities  into spatial Fourier harmonics, 
\begin{equation}\label{eq:Fourier}
f(\bm{r},t)=\int \bar{f}(\bm{k},t){\rm exp}(i\bm{k}\cdot \bm{r})d^3\bm{k}, 
\end{equation}
where $f\equiv (\bm{u}, P, \bm{b})$ and their Fourier transforms are $\bar{f}\equiv (\bar{\bm{u}}, \bar{P}, \bar{\bm{b}})$. Assuming perturbations to be small, linearizing Eqs. (\ref{eq:mom})-(\ref{eq:divu}) and using decomposition (\ref{eq:Fourier}), we get
\begin{equation}\label{eq:mom_linear}
\frac{d{\bar{\bm{u}}}}{dt}-S\bar{u}_x\bm{e}_y=-\frac{i{\bm k}(t)}{\rho}\bar{P}+i(\bm{U}_A\cdot\bm{k}(t))\bar{\bm{b}},
\end{equation}
\begin{equation}\label{eq:ind_linear}
    \frac{d\bar{\bm{b}}}{dt}=-S\bar{b}_x\bm{e}_y+i(\bm{U}_A\cdot\bm{k}(t))\bar{\bm{u}},
\end{equation}
\begin{equation}\label{eq:divu_linear}
i\bm{k}(t)\cdot\bar{\bm{u}}=i\bm{k}(t)\cdot\bar{\bm{b}}=0,
\end{equation}
where we have taken into account that the wavenumber $k_x$ of Fourier modes in the shearwise $x$-direction changes linearly with time due to shear, $k_x(t)=k_x(0)+Sk_yt$, which physically means shearing of the spatial harmonics by the base flow \cite{Gogoberidze2004,chagelishvili2016, Gogichaishvili2014}, while the wavenumbers, $k_y$, in the streamwise $y$-direction, and $k_z$ in the vertical $z$-direction are constant in time. As a result, the wavevector $\bm{k}(t)=(k_x(t),k_y,k_z)$ also varies with time and changes direction.  For simplicity, viscous and resistive dissipation terms are omitted in Eqs. (\ref{eq:mom_linear}) and (\ref{eq:ind_linear}),  respectively,  as they are not important for mode classification and dynamics discussed below.

It follows from Eqs. (\ref{eq:mom_linear})-(\ref{eq:divu_linear}) that two main types of linear modes exist in the flow: (i) Alfv\'en waves, which oscillate with frequencies $\omega=\pm U_Ak_y$, each for $\bm{Z}^{\pm}$, and vary in the streamwise $y$-direction, parallel to the background field $\bm{B}_0$, with the nonzero wavenumber $k_y\neq 0$ and (ii) non-oscillatory,  uniform in the streamwise direction ($k_y=0$) modes,  which have independent magnetic and velocity perturbations. The Alfv\'en wave mode can be further divided into pseudo-Alfv\'en and shear-Alfv\'en waves according to their polarizations \cite{Goldreich-Sridhar1995,Maron2001, Gogichaishvili2014}, which in Fourier space take the form:
\begin{itemize}
\item
\textit{Pseudo-Alfv\'en wave} (PAW) with velocity $\bar{\bm{u}}_p=(\bar{u}_{x},-k_x(t)\bar{u}_x/k_y,0)$ and magnetic field  $\bar{\bm{b}}_p=(\bar{b}_{x},-k_x(t)\bar{b}_x/k_y,0)$ perturbations lying in the $(x,y)$-plane and satisfying $k_x(t)\bar{u}_{p,x}+k_y\bar{u}_{p,y}=k_x(t)\bar{b}_{p,x}+k_y\bar{b}_{p,y}=0$. The corresponding Els\"asser variables are $\bar{\bm Z}^{\pm}_p=(\bar{Z}^{\pm}_x, -k_x(t)\bar{Z}^{\pm}_x/k_y, 0)$ and obey $k_x(t)\bar{Z}_{p,x}^{\pm}+k_y\bar{Z}_{p,y}^{\pm}=0$.

\item
\textit{Shear-Alfv\'en wave} (SAW) with velocity $\bar{\bm{u}}_s=(0,-k_z\bar{u}_z/k_y, \bar{u}_{z})$ and magnetic field $\bar{\bm{b}}_s=(0,-k_z\bar{b}_z/k_y,\bar{b}_{z})$ perturbations lying in the $(y,z)$-plane and satisfying $k_y\bar{u}_{s,y}+k_z\bar{u}_{s,z}=k_y\bar{b}_{s,y}+k_z\bar{b}_{s,z}=0$. The corresponding Els\"asser variables are $\bar{\bm{Z}}^{\pm}_s=(0, -k_z\bar{Z}^{\pm}_z/k_y, \bar{Z}^{\pm}_z)$ and obey $k_y\bar{Z}_{s,y}^{\pm}+k_z\bar{Z}_{s,z}^{\pm}=0$.
\end{itemize}
To analyze the dynamics of PAWs and SAWs, their interaction with each other and with the base flow due to shear $S$, we derive linear equations for $\bar{Z}^{\pm}_{p,x}$ and $\bar{Z}^{\pm}_{s,z}$, which are the independent components of the corresponding Els\"asser variables for these waves, using Eqs. (\ref{eq:mom_linear})-(\ref{eq:divu_linear}) or, equivalently linearizing Eq.  (\ref{eq:Elsassers}) and applying Fourier transform Eq. (\ref{eq:Fourier}), 
\begin{equation}\label{eq:PAW}
\left(\frac{d}{dt}\mp ik_yU_A+S\frac{k_x(t)k_y}{k^2(t)} \right)\bar{Z}_{p,x}^{\pm}=-S\frac{k_x(t)k_y}{k^2(t)}\bar{Z}_{p,x}^{\mp},
\end{equation}
\begin{equation}\label{eq:SAW}
\left(\frac{d}{dt}\mp ik_yU_A \right)\bar{Z}_{s,z}^{\pm}=-S\frac{k_yk_z}{k^2(t)}\left(\bar{Z}_{p,x}^{+}+\bar{Z}_{p,x}^{-}\right),
\end{equation}
where $k^2(t)=k_x^2(t)+k_y^2+k_z^2$. In the absence of shear $S=0$, Eqs. (\ref{eq:PAW}) and (\ref{eq:SAW}) decouple and describe the individual dynamics of counter-propagating PAWs and SAWs, respectively. These waves evolve independently from each other and, as mentioned above, oscillate with the same Alfv\'en frequencies, 
\[
\bar{Z}_{p,x}^{\pm}\propto {\rm e}^{\pm ik_yU_At},~~~~\bar{Z}_{s,z}^{\pm}\propto {\rm e}^{\pm ik_yU_At},
\]
The situation qualitatively changes in the presence of shear, which dynamically couples the waves both to one another and to the base flow. Namely, it gives rise to two main effects for PAWs in Eq. (\ref{eq:PAW}): energy exchange with the base flow for each $\bar{Z}_{p,x}^{\pm}$ component through the third term on the left hand side, $Sk_x(t)k_y\bar{Z}_{p,x}^{\pm}/k^2(t)$, and coupling between these two counter-propagating components through the term $-Sk_x(t)k_y\bar{Z}_{p,x}^{\mp}/k^2(t)$ on the right hand side. Physically, this means that  initially imposed one of these two components generates another -- counter-propagating -- one in the course of evolution \cite{Gogoberidze2004, Gogichaishvili2014}. The coefficient $Sk_x(t)k_y/k^2(t)$ which is responsible for these two shear effects is appreciable at $|k_x(t)|\lesssim |k_y|$, during the shear time $\sim S^{-1}$, and vanishes at large wavenumber $k(t)$. Therefore, for the shear-induced non-modal growth and coupling of Alfv\'en waves to be effective, the wave period should be larger than the shear time, i.e., the wave frequency should be smaller than the shear, $k_yU_A\lesssim S$ \cite{Gogichaishvili2014,mamatsashvili2014}.
Note that SAWs do not enter Eq. (\ref{eq:PAW}) and hence do not affect the dynamics of PAWs in the linear regime. These two shear-induced effects jointly give rise to the non-modal growth of both $\bar{Z}_{p,x}^{\pm}$ branches of PAWs and their over-reflection.

To clearly demonstrate this over-reflection phenomenon and its relation to the turbulence balancing, we use the wave action conservation for PAWs \cite{Gogoberidze2004,Gogichaishvili2014}.  Multiplying Eq. (\ref{eq:PAW}) by the complex conjugate $(\bar{Z}_{p,x}^{\pm})^{\ast}$,  adding to its associated conjugate equation and then subtracting the equation for $|\bar{Z}_{p,x}^{-}|^2$ from equation for $|\bar{Z}_{p,x}^{+}|^2$, we get
\begin{equation}\label{eq:conservation}
k^2(t)(|\bar{Z}_{p,x}^{+}|^2-|\bar{Z}_{p,x}^{-}|^2)=const.
\end{equation}
Suppose that only $\bar{Z}_{p,x}^{+}(0)\neq 0$ is present initially in the flow, which we refer to as the incident wave, while $\bar{Z}_{p,x}^{-}(0)=0$, i.e., the initial state is perfectly imbalanced.  As discussed above, the non-modal growth and coupling of waves occur during the time interval in which $|k_x(t)|\lesssim |k_y|$, resulting in the generation and amplification of the other -- counter-propagating -- component $\bar{Z}_{p,x}^{-}$ by the initially imposed one $\bar{Z}_{p,x}^{+}$. We apply Eq. (\ref{eq:conservation}) at the initial moment $t=0$, when $k_x(0)/k_y<0$ and there is only $\bar{Z}_{p,x}^{+}\neq 0$, and at a final moment $t_f$, after the growth and coupling processes have taken place, when $k_x(t_f)/k_y>0$ and there are both $\bar{Z}_{p,x}^{+}$ and newly generated $\bar{Z}_{p,x}^{-}$. Choosing the initial and final moments such that $k_x(0)=-k_x(t_f)$, we get,
\begin{equation}\label{eq:conservation1}
|\bar{Z}_{p,x}^{+}(0)|^2=|\bar{Z}_{p,x}^{+}(t_f)|^2-|\bar{Z}_{p,x}^{-}(t_f)|^2.
\end{equation}
This relation indicates that the amplitude of $\bar{Z}_{p,x}^{+}$ is higher after the non-modal growth than its initial amplitude. The counter-propagating component $\bar{Z}_{p,x}^{-}$ is referred to as the reflected wave,  while the $\bar{Z}_{p,x}^{+}$ becomes a transmitted wave after the coupling process. It was shown in \cite{Gogichaishvili2014, mamatsashvili2014} that PAWs undergo a stronger non-modal growth, $|\bar{Z}_{p,x}^{\pm}(0)| \ll |\bar{Z}_{p,x}^{\pm}(t_f)|$,  at those $k_y$ for which  the shear effects are important, $k_yU_A\lesssim S$, therefore it follows from Eq. (\ref{eq:conservation1}) that the amplitudes of both transmitted and reflected waves are nearly equal and larger than the amplitude of the incident wave, $|\bar{Z}_{p,x}^{+}(t_f)| \approx |\bar{Z}_{p,x}^{-}(t_f)|\gg |\bar{Z}_{p,x}^{+}(0)|$ and therefore this process is referred to as wave over-reflection. Energy for both PAW components is extracted from the base flow via the non-modal growth process due to shear, making over-reflection possible. The shear-induced over-reflection can thus bring an initially imbalanced state toward balance, a process that, as we demonstrate in the next section, also occurs in shear-driven MHD turbulence composed of Alfv\'en waves. The degree of Alfv\'en wave balance increases as the ratio $k_yU_A/S$ decreases \cite{Gogichaishvili2014}.
 
 As for SAWs, their dynamics is governed by Eq. (\ref{eq:SAW}), from which it follows that these waves are coupled only to PAWs. In other words, SAWs  gain energy from PAWs solely through the shear-induced coupling term on the right-hand side of this equation, without directly exchanging energy with the base flow.

Thus, PAWs play an important dynamical role in the presence of shear. The non-modal growth and over-reflection processes they exhibit, together with their shear-induced coupling to SAWs, necessitate accounting for PAWs in the dynamical balances of shear-driven MHD turbulence. By contrast,  in shearless MHD turbulence driven by external forcing, SAWs are generally assumed to play a dominant role, whereas PAWs are usually neglected \cite{Goldreich-Sridhar1995, Mason_etal2006,beresnyak-2010, boldyrev-2009}.

 \begin{table}
\setlength{\tabcolsep}{6pt}
\renewcommand{\arraystretch}{1.15}
\begin{tabular}{c c c c c}
\toprule
{$M_A$} &
{\(\langle\langle E_K\rangle\rangle\)} &
{\(\langle\langle E_M\rangle\rangle\)} &
{\(\langle \langle u_xu_y\rangle\rangle\)} &
{\(\langle \langle -b_xb_y\rangle\rangle\)} \\
\hline
\(\infty\) &  1.53 & \text{--}& 0.4 & \text{--}\\
\(10\) &  0.84 & 0.24 & 0.15 & 0.023 \\
\(5\) &  0.74 & 0.21 & 0.12 & 0.019 \\
\(3.3\) & 0.68 & 0.19 & 0.11 & 0.018 \\
\hline
\end{tabular}
\caption{Simulation parameters:  Alfv\'en Mach number $M_A$, including the hydrodynamic case $M_A=\infty$, and the volume- and time-averaged (denoted by double brackets) kinetic, $E_{kin}$, and  magnetic, $E_{mag}$, energy densities as well as the Reynolds, $u_x u_y$, and Maxwell, $ -b_x b_y$, stresses in the turbulent state.}
\label{tab:runs}
\end{table}

\section{Simulation Results}

Having described the linear dynamics of  Alfv\'en waves,  we now investigate their nonlinear evolution in shear-driven MHD turbulence by solving the governing Eqs. (\ref{eq:mom})-(\ref{eq:divu}) in a cubic domain with sizes $(L_{x}, L_{y}, L_{z})=(2\pi, 2\pi, 2\pi)\ell$ using the pseudo-spectral code SNOOPY \cite{Lesur2007},  where $\ell$ is some arbitrary length scale of the order of the flow system size.  As is customary in simulations of constant shear flows without rigid boundaries, the boundary conditions are periodic in $y$ and $z$, but shear-periodic in $x$ (e.g., \cite{Hawley_etal1995, Umurhan2004, Lesur2005}).  In this domain, the wavenumbers along the $y$- and $z$-directions are discrete, $k_y=2\pi n_y/L_y$ and $k_z=2\pi n_z/L_z$, where the integer $n_y,n_z=0, \pm1, \pm2, \pm3,...$, while the wavenumber in the $x$-direction, as mentioned above, changes with time as $k_x(t)=2\pi n_x/L_x+St(2\pi n_y/L_y)$ due to shear  for $n_y\neq 0$ to satisfy the shearing-periodic boundary conditions \cite{Hawley_etal1995}, where $n_x=0,\pm1, \pm2, \pm3,...$. $k_x$ coincides with the discrete values $2\pi n_x/L_x$ at specific moments $t_n=nL_y/(SL_xn_y)$, where $n$ is a positive integer.

We non-dimensionalize time by $S^{-1}$, length by $\ell$, velocity by $S\ell$ (magnetic field is measured in velocity units too) and energy densities by $\rho_0S^{2}\ell^{2}$. The main parameters of the problem are the Reynolds number $Re=S\ell^2/\nu$, magnetic Reynolds number $Rm=S\ell^2/\eta$ and Alfv\'en Mach number $M_A=S\ell/U_{A}=(4\pi\rho)^{1/2} S\ell/B_{0y}$, which characterizes the effect of the shear relative to the Lorentz force due to the background field. We fix $Re = Rm=1000$, so that the magnetic Prandtl number $Pm=Rm/Re=1$. As noted above, the shear-induced linear processes -- non-modal growth and coupling -- of Alfv\'en waves are efficient when their frequency is lower than the shear, $k_yU_A\lesssim S$, or equivalently, $k_y\ell \lesssim M_A$. This efficiency increases with decreasing ratio $k_yU_A/S$, i.e., with increasing shear \cite{Gogichaishvili2014}. For this condition to be satisfied  at least for the smallest $k_{y, min}=2\pi/L_y$ in the domain, it is necessary that $M_A \gtrsim 2\pi\ell/L_y$, or $M_A \gtrsim 1$, which implies the \textit{super-Alfv\'enic regime}. Therefore, the following analysis focuses mainly on the case $M_A=10$, representative of the strongly super-Alfv\'enic regime $M_A\gg 1$ typically observed in the near-Earth solar wind \cite{Jian2011, Borovsky2019}. For comparison, we also did simulations for lower $M_A=3.3$ and 5 as well as for the purely hydrodynamic case (i.e., $\bm{b}\equiv 0$ and $M_A=\infty$), which are all listed in Table \ref{tab:runs}.

The numerical resolution of the simulations is $(N_{x}, N_{y}, N_{z})=(256, 256, 256)$. To verify that this resolution is sufficient for the present problem, we compare viscous, $\lambda_{\nu}= (\nu^3/\epsilon)^{1/4}$, and resistive, $\lambda_{\eta}=Pm^{-3/4}\lambda_{\nu}$, dissipation scales of MHD turbulence \cite{Rincon2019}, which are equal, $\lambda_{\nu}=\lambda_{\eta}$, for $Pm=1$, with the largest resolved wavenumber in the domain $k_{max}=2\pi N_x/3L_x$ accessible by the pseudo-spectral code (using the standard 2/3 dealiasing rule). Here $\epsilon=\nu\langle(\nabla \times {\bm u})^2\rangle$ is the volume-averaged dissipation rate of the kinetic energy. For our simulations, we obtain $k_{\max}\lambda_{\nu}=k_{\max}\lambda_{\eta}\approx 1$, indicating that both dissipation scales are resolved at the standard spectral resolution threshold \cite{Ishihara2007}. Thus, the adopted resolution is adequate, since the dominant wave dynamics in fact occurs at wavenumbers lower than $k_{max}$, as shown by the spectral analysis in Sec. III C.
As an additional check, we also carried out simulations for $M_A=10$ by doubling the  resolution, $(N_x,N_y,N_z)=(512,512,512)$. The results show good qualitative and quantitative agreement in terms of the time-averaged energies and stresses with those obtained at $(N_x,N_y,N_z)=(256,256,256)$, demonstrating convergence with resolution.

\begin{figure}[t!]
  \centering
  \includegraphics[width=0.5\textwidth]{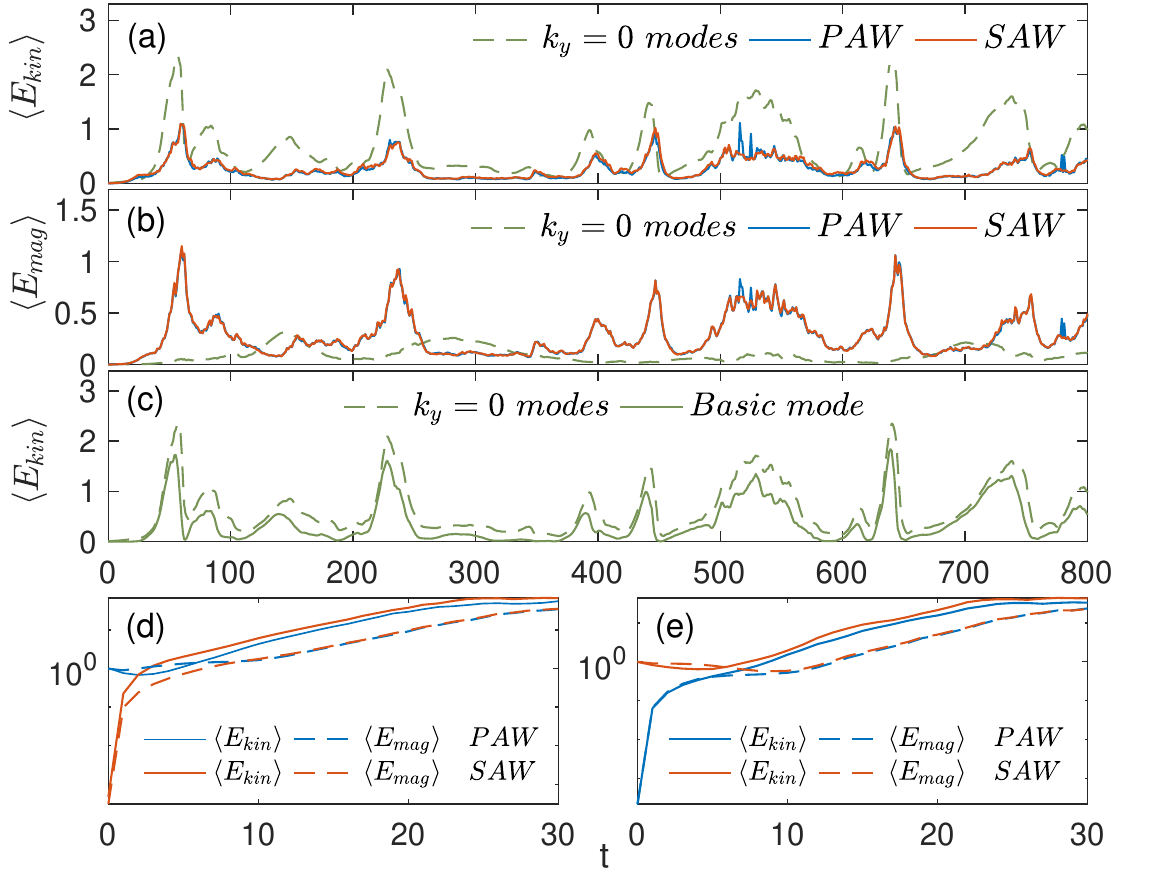} 
  \caption{Evolution of the volume-averaged (a) kinetic $\langle E_{kin} \rangle$ and (b) magnetic $\langle E_{mag}\rangle$ energies of PAW, SAW and $k_{y}=0$ modes for the initially imposed PAWs at $M_A=10$. (c) Kinetic energies of the basic mode and all of the $k_y=0$ modes. (d) The same evolution of the wave energies as in (a) and (b), but at early times and normalized by the initial values. (e) The same as in (d) but for initially imposed SAWs.} \label{fig:1}
\end{figure}
\begin{figure*}[t!]
  \includegraphics[scale=0.3]{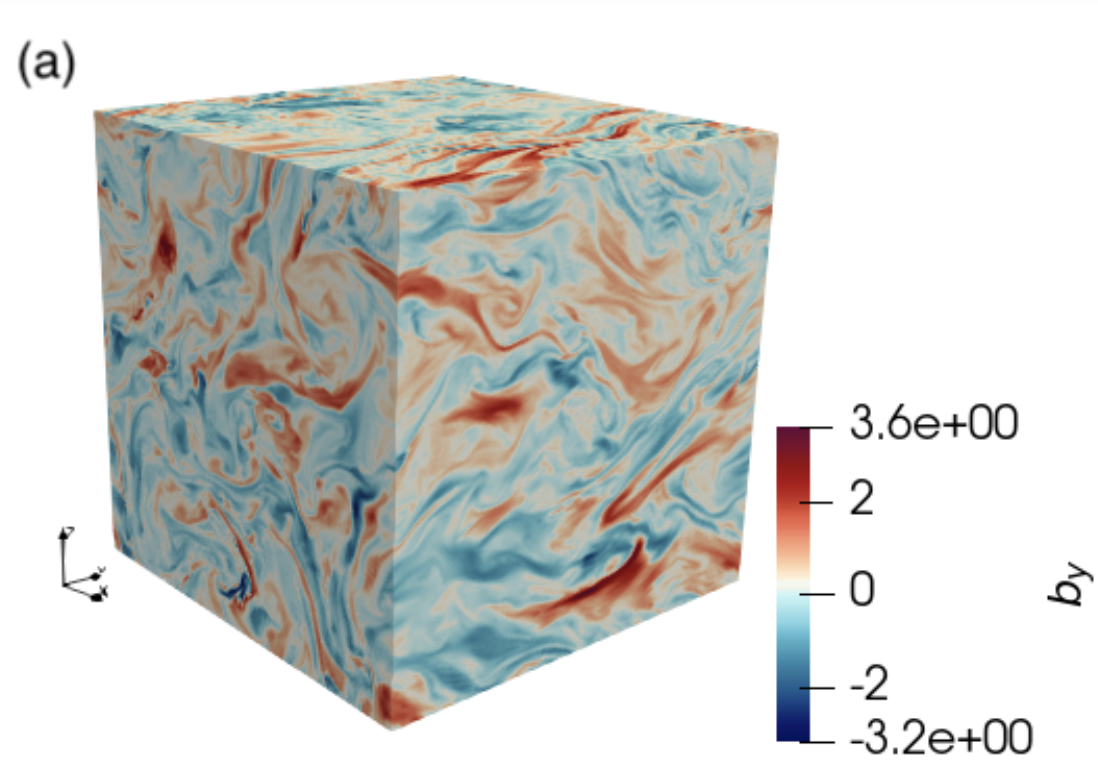} 
  \includegraphics[scale=0.3]{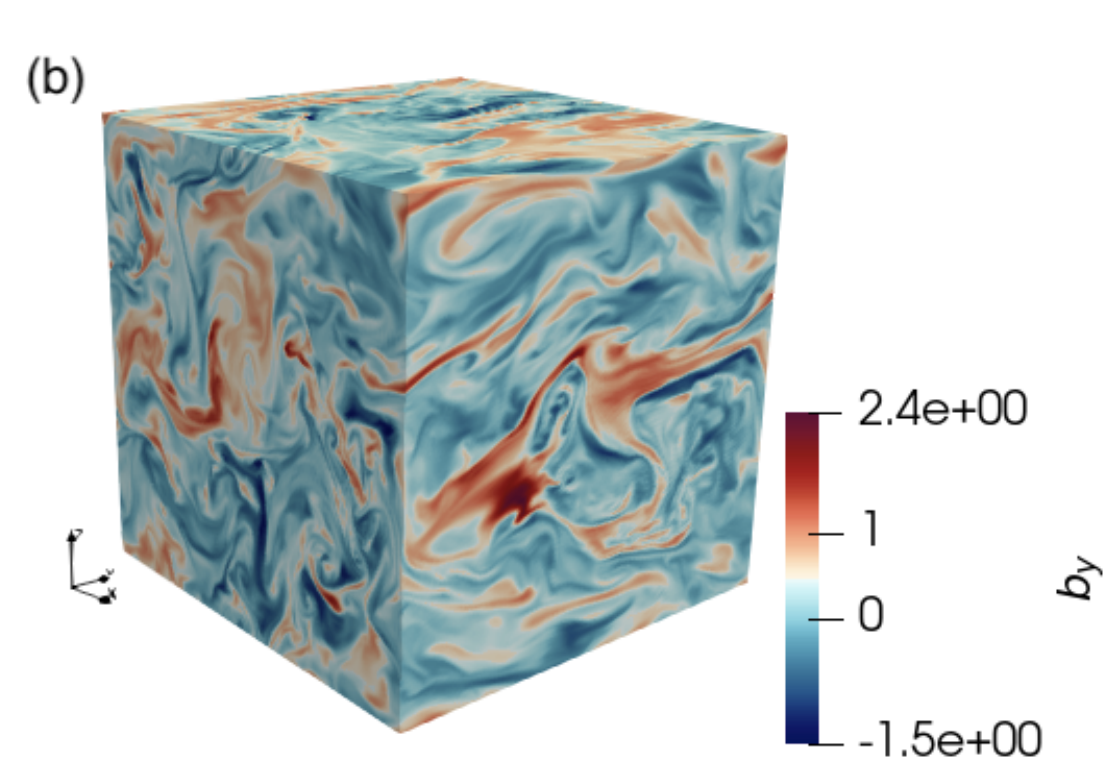}
  \includegraphics[scale=0.3]{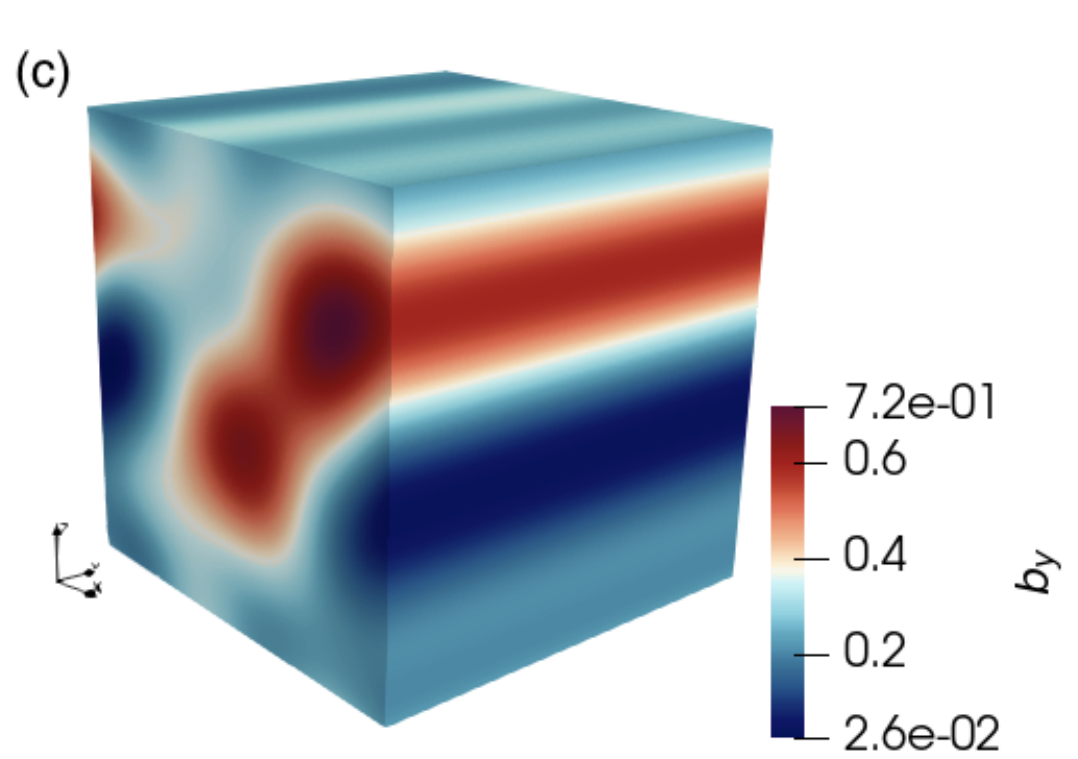}
  \caption{Structure of the streamwise field component,  $b_y$,  in physical space in the fully developed MHD turbulence at (a) $M_A=10$ and (b) $M_A=5$ as well as in the laminar state at (c) $M_A=2.94$. The laminar state is characterized by uniform  in the streamwise $y$-direction large-scale modes. All these snapshots are taken at $t=670$.} \label{fig:2}
\end{figure*}

Since we are mainly interested in the role of shear in the balancing of MHD turbulence, we consider two types of perfectly imbalanced initial conditions composed of random perturbations of: (i) only PAWs with nonzero $\langle(Z_p^{+})^2\rangle^{1/2}=0.14$ and zero $Z_p^{-}=Z_s^{\pm}=0$ (i.e., no SAWs) and (ii) only SAWs with nonzero $\langle(Z_s^{+})^2\rangle^{1/2}=0.16$ and zero $Z_s^{-}=Z_p^{\pm}=0$ (i.e., no PAWs). The corresponding rms amplitudes of initial velocity and magnetic field perturbations are, respectively,  $\langle u^2\rangle^{1/2}=\langle b^2\rangle^{1/2}=0.1$ in the first and $\langle u^2\rangle^{1/2}=\langle b^2\rangle^{1/2}=0.11$ in the second type of the initial conditions. The $k_y=0$ non-oscillatory modes are initially absent in both cases. Other imbalanced initial conditions with $Z_{p,s}^{-}\neq 0$ and $Z_{p,s}^{+}=0$ exhibit qualitatively similar evolution (not shown here).

The initial perturbations undergo non-modal amplification due to shear and after several shear times their amplitude becomes sufficiently high to reach a nonlinear regime.  As a result, the flow eventually settles into a quasi-steady MHD turbulence with characteristic bursts in the kinetic and magnetic energies (Figs. \ref{fig:1} and \ref{fig:2}).  Because the base shear flow is modally stable, it supports no exponentially growing eigenmodes but only non-modally growing perturbations and, consequently, the transition to turbulence is subcritical \cite{Schmid-Henningson2001}. This means that, in addition to the main parameters of the flow ($Re$, $Rm$, $M_A$), the onset of turbulence also depends on the initial amplitude of the perturbations: turbulence sets in and persists only when this amplitude exceeds a certain threshold. In this case, non-modal growth of perturbations is considered one of the key mechanism for triggering transition to turbulence in subcritical regimes in shear flows \cite{Gebhardt_Grossmann1994,Baggett_etal1995, Grossmann2000}. Here, however, we do not address this transition process itself, that is, how the critical amplitude of the initial perturbations depends on the flow parameters and the domain aspect ratios ($L_z/L_x$, $L_y/L_x$), but instead focus on the dynamics and balancing of fully developed MHD turbulence driven by shear. For this purpose, we take initial velocity and magnetic field perturbations with sufficiently high amplitude to ensure that the flow always exhibits transition to sustained turbulence for the given other parameters.

\subsection{Shear-driven MHD turbulence -- nonlinear evolution of PAWs, SAWs and the $k_y=0$ modes}

First, we describe the general properties of the turbulence and then its balanced/imbalanced dynamics. Figure \ref{fig:1} shows the evolution of the volume-averaged kinetic, $\langle E_{kin} \rangle=\langle u^2\rangle/2$, and magnetic, $\langle E_{mag} \rangle=\langle b^2\rangle/2$, energy densities of PAWs and SAWs with $k_y\neq 0$ as well as the streamwise-uniform $k_y=0$ modes,  starting from the perfectly imbalanced states described above.  Figures \ref{fig:2}(a) and \ref{fig:2}(b) show the typical structure of this fully developed MHD turbulence through the distribution of the dominant streamwise field component, $b_y$, in physical space. The field is filled with chaotic small-scale structures stretched along the $y$-direction by the shear.  Note that the kinetic and magnetic energies of PAWs and SAWs each are almost equal at all times  [Figs. \ref{fig:1}(a) and \ref{fig:1}(b)],  that is,  there is equipartition between the PAW and SAW energies in contrast to MHD turbulence without shear, where, as noted above, only SAWs play a role.  

The $k_{y}=0$ modes, which have been initially absent, emerge as a result of nonlinear interaction of Alfv\'en waves during the early growth stage and in the subsequent turbulent state undergo occasional rapid non-modal growth events,  or bursts [Fig.  \ref{fig:1}(c)]. These bursts are in fact associated with the bursts of the dominant ${\bm k}_b=(0,0,2\pi/L_z)$ mode, referred to as the \textit{basic mode},  since it carries the highest kinetic energy among all the $k_y=0$ modes.  During each burst, the basic mode first rapidly grows and then decays, nonlinearly transferring energy to the Alfv\'en waves that results in similar bursts in the wave energies with a slight time lag [Fig.  \ref{fig:1}(a)].  In contrast to the waves,  the $k_y=0$ modes remain essentially hydrodynamic, because their magnetic energy is much smaller than the kinetic one and smaller than the magnetic energy of the waves. Therefore, this bursty dynamics of the basic mode, and so  all the $k_y=0$ modes,   is governed by the interplay of the mostly hydrodynamic linear non-modal growth and nonlinear processes  \cite{Pumir1996, mamatsashvili2016}.

Although the evolution of the PAW and SAW energies is quite similar in the turbulent state,  their early-time development for the considered both perfectly imbalanced initial states with only PAWs or SAWs  differs due to the essentially different linear dynamics of these waves, as discussed in Sec. II A.  In the first case, initially imposed PAWs undergo non-modal growth (over-reflection) and also cause SAWs to rapidly grow due to linear coupling [Fig. \ref{fig:1}(d)]. By contrast, in the second case,  the growth of the imposed SAWs is slower -- they initially excite PAWs due to nonlinearity, which in turn non-modally grow and reinforce the originally imposed SAWs [Fig. \ref{fig:1}(e)]. 

The time- and volume-averages of the kinetic and magnetic energies of the perturbations as well as the Reynolds, $u_xu_y$, and Maxwell, $-b_xb_y$, stresses for different $M_A$ in the turbulent state are listed in Table \ref{tab:runs}. Note that the kinetic energy and Reynolds stress are, respectively, about 3.5 and 6 times larger than the magnetic energy and Maxwell stress, implying that the turbulence is driven mainly by shear-induced hydrodynamic non-modal processes \cite{Pumir1996, mamatsashvili2016}, but is modified by the magnetic field. Specifically, in the hydrodynamic case ($M_A=\infty$), where the flow is not affected by the field, the kinetic energy and Reynolds stress reach their maximum values and decrease with decreasing $M_A$, i.e., with increasing field strength. The magnetic energy and Maxwell stress, which are mainly associated with Alfv\'en waves, exhibit a similar decreasing trend with $M_A$. This indicates that the background field has a stabilizing effect on the flow: it undergoes a transition from a turbulent to a laminar state at some critical value $M_{A,c}$ as this parameter decreases.  For the given initial conditions,  such a relaminarization occurs at $M_{A,c}\approx 3$.  This laminar state is composed of several large-scale non-oscillatory $k_y=0$ modes that are uniform in the streamwise $y$-direction [Fig. \ref{fig:2}(c)].

\subsection{Balancing of the turbulence}

We now focus on the balance/imbalance properties of the resulting shear-driven MHD turbulence. To this end, we examine the nonlinear evolution of the Els\"asser variables, whose linear dynamics under the influence of shear was described in Sec. II A. 

In analogy with the normalized cross-helicity $\sigma_c$ (Eq. \ref{eq:sigma}), which characterizes an overall imbalance of MHD turbulence, we also define the ratios $f_p$ and $f_s$,
\begin{equation}\label{eq:fp}
f_{p}=\frac{\left\langle (Z_{p}^{+})^{2}\rangle -\langle(Z_{p}^{-})^{2}\right\rangle}{\left\langle (Z_{p}^{+})^{2}\rangle +\langle(Z_{p}^{-})^{2}\right\rangle},
\end{equation}
\begin{equation}\label{eq:fs}
f_{s}=\frac{\left\langle (Z_{s}^{+})^{2}\rangle -\langle(Z_{s}^{-})^{2}\right\rangle}{\left\langle (Z_{s}^{+})^{2}\rangle +\langle(Z_{s}^{-})^{2}\right\rangle},
\end{equation}
which measure the degree of imbalance separately for PAWs and SAWs, respectively, in the turbulent state. Since the two types of initial conditions described above lead to qualitatively similar saturated turbulent states, as evident in Fig.~\ref{fig:1}, we focus on the time evolution of the Els\"asser energies for the first type of the perfectly imbalanced initial condition, which contains only PAWs with nonzero $\bm{Z}_p^{+}\neq 0$ and no SAWs ($\bm{Z}_s^{\pm}=0$). Therefore, initially $f_p(0)=1$, whereas $f_s(0)$ is formally undefined because SAWs are absent. However, SAWs are generated by PAWs shortly after the start [Fig.~\ref{fig:1}(d)], so that $f_s$ becomes well defined already near $t=0$.

\begin{figure}[t!]
\center{\includegraphics[width=0.49\textwidth]{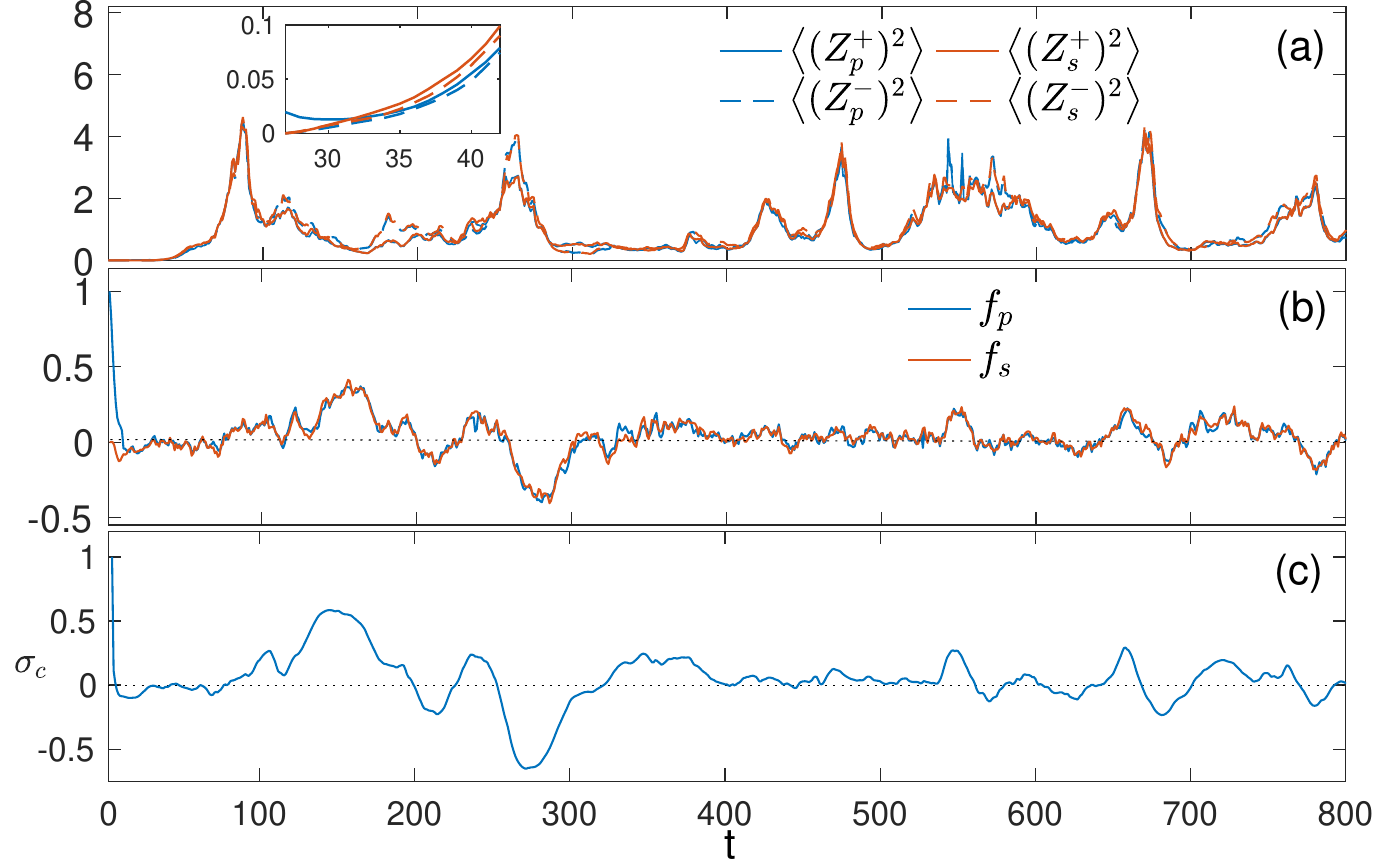}}
  \caption{Evolution of (a) the volume-averaged Els\"asser energies $\langle (Z_p^{\pm})^2\rangle$ and $\langle (Z_s^{\pm})^2 \rangle$ (the factor $1/4$ are omitted for simplicity) for PAWs (blue) and SAWs (red), respectively, for the initially imposed only PAWs at $M_A=10$,  (b) the corresponding ratios $f_{p}$ and $f_s$ [given by Eqs.  (\ref{eq:fp}) and (\ref{eq:fs}),  respectively] and (c) the normalized cross-helicity $\sigma_c$ [Eq. (\ref{eq:sigma})].} \label{fig:3}
\end{figure}

\begin{figure}[t!]
\center{\includegraphics[width=0.49\textwidth]{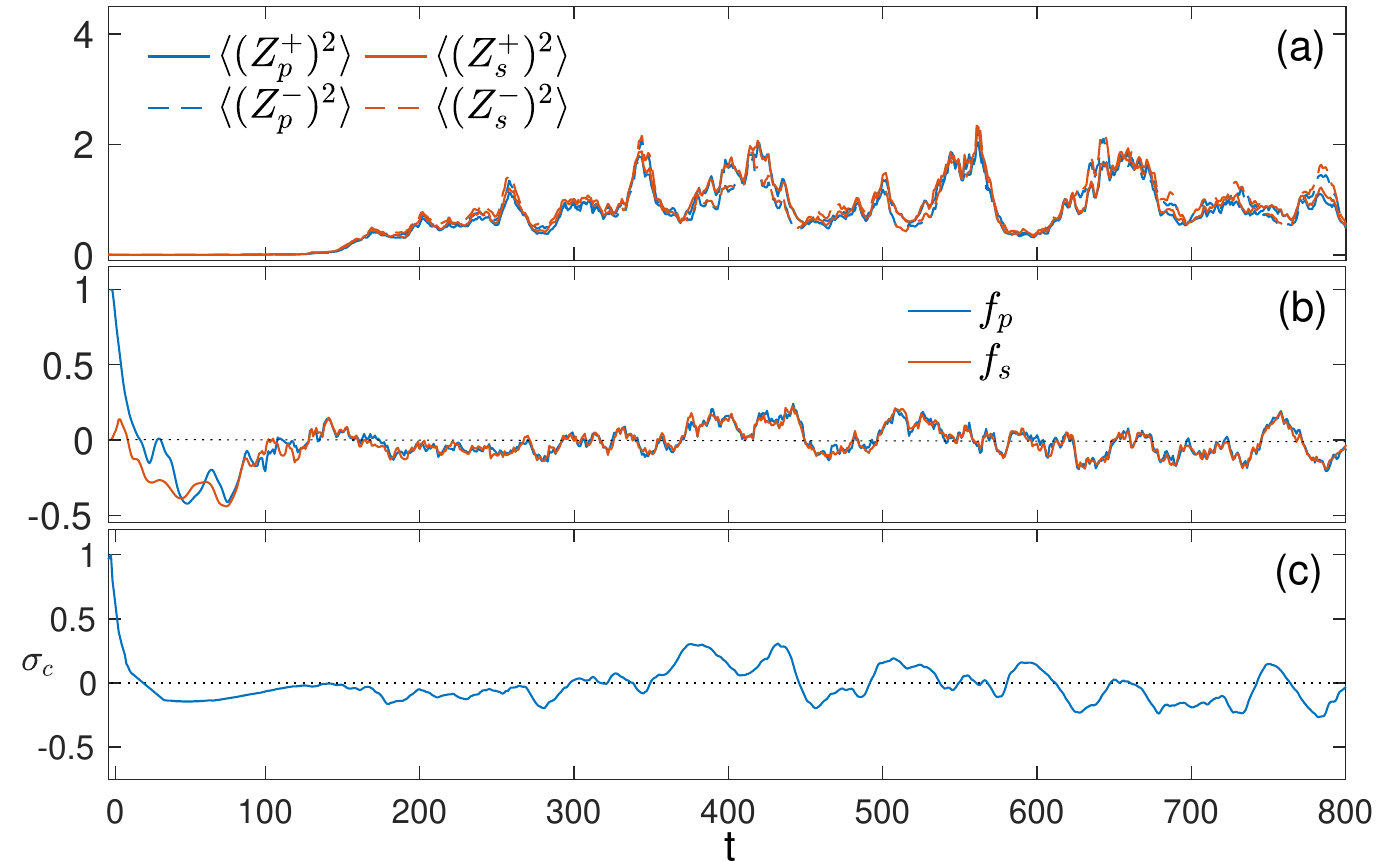}}
  \caption{Same as in Fig. \ref{fig:3} but for $M_A=3.3$.} \label{fig:4}
\end{figure}

Figure \ref{fig:3} shows the evolution of the volume-averaged Els\"asser energies of PAWs and SAWs, the corresponding ratios $f_p$, $f_s$ and $\sigma_c$. It is clear that the initially imposed only PAWs with $\bm{Z}_{p}^{+}\neq 0$, its counterpart $\bm{Z}_{p}^{-}$ generated as a result of non-modal growth (over-reflection) and $\bm{Z}_{s}^{\pm}$ of SAWs excited due to linear coupling with the PAWs all become remarkably close to each other in the turbulent state: $\langle (Z_{p}^{+})^2\rangle \approx \langle (Z_{p}^{-})^2\rangle \approx \langle (Z_{s}^{+})^2\rangle \approx \langle (Z_{s}^{-})^2\rangle$,  as seen in Fig.  \ref{fig:3}(a).  Thus, there is equipartition not only between the Els\"asser energies of PAWs and SAWs (Fig. \ref{fig:1}), but also between $\bm{Z}^{+}$ and $\bm{Z}^{-}$ for each of these Alfv\'en wave branches.  Consequently, $f_p$ rapidly drops from its initial value of 1 and fluctuates near zero in the turbulent state, while $f_s$ closely follows that of $f_p$, remaining approximately equal to it at all times [Fig. \ref{fig:3}(b)]. The evolution of the normalized cross-helicity $\sigma_c$  exhibits a trend similar to that of $f_p$ [Fig. \ref{fig:3}(c)].   The corresponding time-averaged values of $f_p$, $f_s$ and $\sigma_c$ are close to 0, therefore the rms amplitudes of their fluctuations provide a more relevant measure of the degree of turbulence balance.  We find that the rms amplitudes of  these quantities computed over simulation time, $\langle f_p^2\rangle_t^{1/2}=\langle f_s^2\rangle_t^{1/2}=0.08$ and $\langle \sigma_c^2\rangle_t^{1/2}=0.1$, remain much smaller than 1. This indicates that the turbulence is balanced on average in time, despite being initially maximally imbalanced.

The evolution of the Els\"asser energies as well as $f_p$, $f_s$ and $\sigma_c$ towards balanced MHD turbulence also occurs at lower $M_A$, as shown in Fig. \ref{fig:4} for $M_A=3.3$,  which is qualitatively similar to that for $M_A=10$ in Fig. \ref{fig:3}.  The time-averaged Els\"asser energies are all close to one another, but are about 1.3 times smaller than those at $M_A=10$; their peak values are also lower. Being in the balanced state, $f_p$, $f_s$ and $\sigma_c$ fluctuate similarly around 0, however, their rms amplitudes, $\langle f_p^2\rangle_t^{1/2}=\langle f_s^2\rangle_t^{1/2}=0.1$ and $\langle \sigma_c^2\rangle_t^{1/2}=0.13$, are  slightly larger than those at $M_A=10$, implying that the degree of turbulence balance improves with increasing $M_A$.

\begin{figure}[t!]
  \centering
  \includegraphics[scale=0.11]{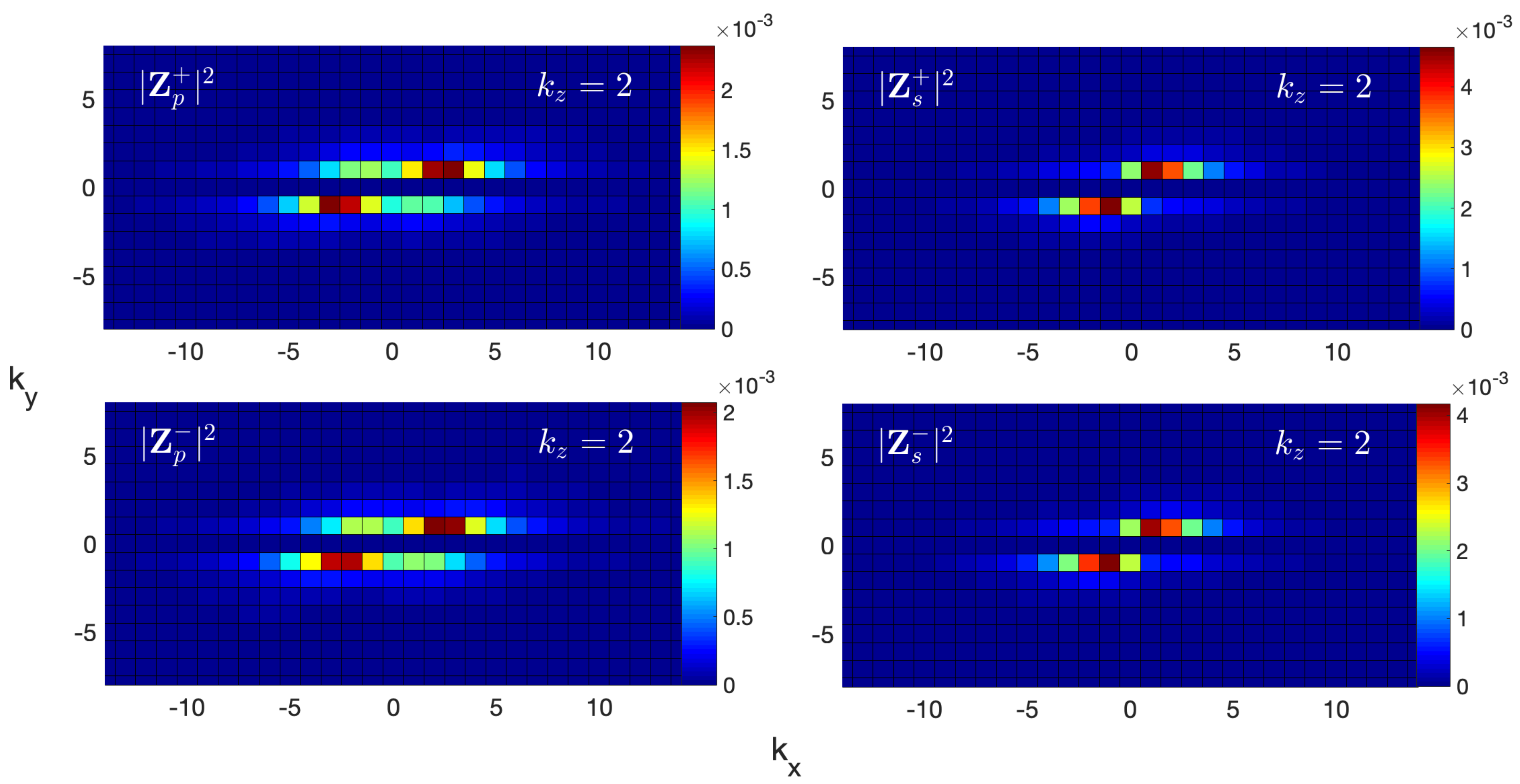}
\vspace{-1.5\baselineskip}
  \caption{The $(k_x, k_y)$-section of the 3D spectra of $|\bar{\bm{Z}}^{\pm}_p|^2$ (left) and $|\bar{\bm Z}^{\pm}_s|^2$  (right) at $k_z =2$ in the turbulent state at $M_A=10$. } \label{fig:5}
\end{figure}

\begin{figure*}[t!]
  \centering
  \includegraphics[width=\textwidth]{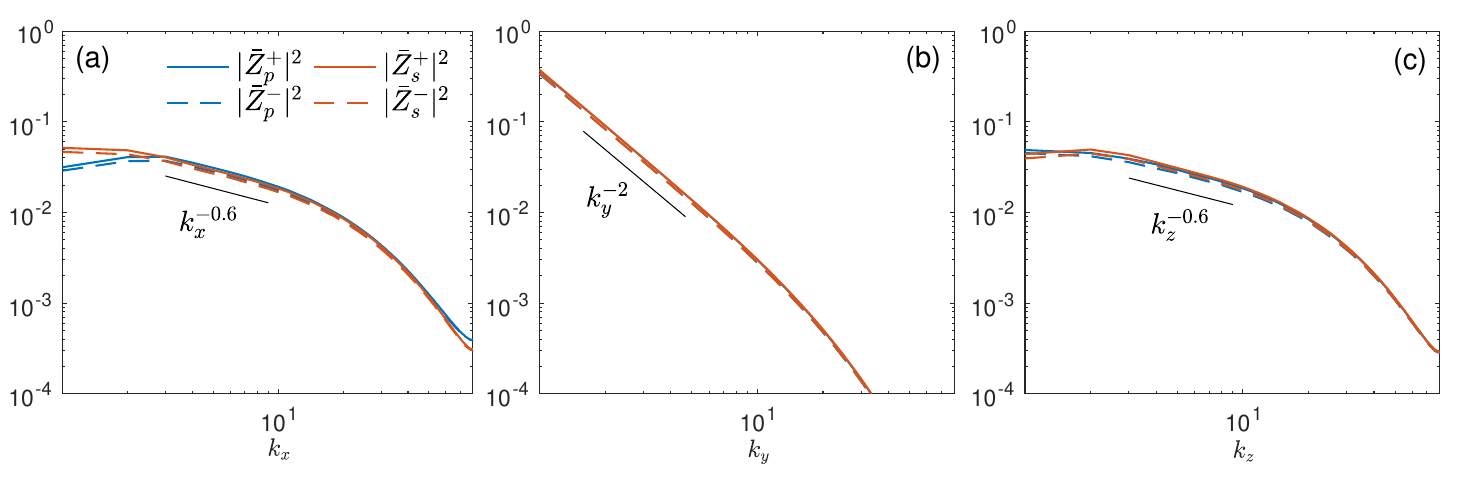}
\vspace{-2.\baselineskip}
  \caption{1D spectra of $|\bar{\bm Z}^{\pm}_p|^2$ and $|\bar{\bm Z}^{\pm}_s|^2$ represented as a function of (a) $k_x$, (b) $k_y$ and (c) $k_z$ being integrated in each case over the other two wavenumbers. }\label{fig:6}
\end{figure*}

This efficient exchange among all four Alfv\'en wave branches --  counter-propagating two PAWs ($\bm{Z}_p^{\pm}$) and two SAWs ($\bm{Z}_s^{\pm}$) -- which results in the emergence of this  balanced state from an initially imbalanced one, is therefore due to shear-induced non-modal dynamics of Alfv\'en waves, described in Sec. II A. Indeed, in the classical case of shearless MHD turbulence, Els\"asser fields interact only through nonlinear processes that do not change their respective total energies $(1/4)\int(Z_{p,s}^{+})^2dV$ and $(1/4)\int(Z_{p,s}^{-})^2dV$,  but merely  redistribute these energies among different modes such that the degree of turbulence imbalance remains  conserved in time \cite{Goldreich-Sridhar1995}.

\subsection{Spectral analysis}

We also examine the energy spectrum of Alfv\'en waves, which is one of the key characteristics of MHD turbulence.  Figure \ref{fig:5} shows the $(k_x,k_y)$-section of the time-averaged 3D spectra of all four Els\"asser fields $\bar{\bm{Z}}^{\pm}_p$ and  $\bar{\bm{Z}}^{\pm}_s$ in Fourier space at a given $k_z=2$ where they reach high values. These spectra appear similar because the turbulence is on average balanced and exhibit pronounced anisotropy due to shear with more power being concentrated in the $k_xk_y>0$ region than in the $k_xk_y<0$ region, as also observed in other studies of shear-driven MHD turbulence  \cite{mamatsashvili2014,  Gogichaishvili-2017, Murphy2015, Mamatsashvili2020, held2022}. This shear-induced spectral anisotropy arises because  individual Alfv\'en waves undergo linear non-modal growth as their time-varying $k_x(t)$ drifts in Fourier space from the region where $k_xk_y<0$ to $k_xk_y>0$, as discussed in Sec. II A.  It is topologically distinct from the well-known  spectral anisotropy of forced  MHD turbulence without a mean flow shear, which is caused by a background (guide) magnetic field (e.g., \cite{Goldreich-Sridhar1995, cho-2000, Maron2001, Boldyrev_2006PhRvL, boldyrev-2009}): the former occurs between the streamwise and shearwise directions of the base flow, while the latter occurs between the directions parallel and perpendicular to the background field.

Figure \ref{fig:6} shows the 1D spectra of the Els\"asser fields in the perpendicular ($k_x$ and $k_z$) and parallel ($k_y$) directions to the base flow and background field, which are obtained by integrating their 3D spectra over the other two directions. These 1D spectra also appear to be identical for both PAWs and SAWs,  implying that balance is achieved in fact over a broad range of scales. They differ most notably along $k_x$ and $k_y$, with the latter being somewhat steeper than the former, which is a consequence of the anisotropy of the 3D spectra in the $(k_x,k_y)$-plane (Fig. \ref{fig:5}).  On the other hand, the 1D spectra along $k_x$ and $k_z$ are mostly similar with only a slight difference at low wavenumbers. The $k_y$-spectra follow $k_y^{-2}$ as is the case in forced MHD turbulence  \cite{Beresnyak2015}. However, the spectra along $k_x$ and $k_z$,  which follow the same power-laws $k_x^{-0.6}$ and $k_z^{-0.6}$ at intermediate wavenumbers,  differ from the classical ones $k^{-5/3}$ or $k^{-3/2}$ \cite{Goldreich-Sridhar1995, lithwick-2007, Boldyrev_2006PhRvL, perez-2012} and this difference can be attributed to shear.  This again confirms that the spectral anisotropy of MHD turbulence resulting from shear is fundamentally different from that in the classical shearless case.

\section{Conclusions}

In this paper, we studied the dynamics of MHD turbulence in shear flows with a uniform streamwise background magnetic field in the super-Alfv\'enic regime when Alfv\'en Mach number $M_A\gtrsim 1$. We focused on the effect of large-scale shear of the flow velocity on the balanced/imbalanced nature of the turbulence. In this regime, shear plays an important role in the dynamics of Alfv\'en waves with frequencies smaller than the shear rate and thus affects the overall dynamics of MHD turbulence composed of interacting Alfv\'en waves. We demonstrated the reduction of turbulence imbalance in this regime, showing that for sufficiently strong shear -- already at $M_A \approx  3$ and higher, i.e., in the strongly super-Alfv\'enic regime $M_A\gg 1$ -- the turbulence becomes essentially balanced. By contrast, in classical forced MHD turbulence in the absence of mean flow shear,  the degree of imbalance is conserved.  We showed that the dynamics of pseudo-Alfv\'en and shear-Alfv\'en waves, which are the building blocks of MHD turbulence,  is governed by shear-induced linear processes -- non-modal growth, manifested as the wave over-reflection process, and mode coupling. The non-modal growth of pseudo-Alfv\'en waves supplies energy to turbulence, whereas the linear  coupling leads to energy exchange and equipartition between pseudo-Alfv\'en and shear-Alfv\'en waves as well as between the counter-propagating components $\bm{Z}^{\pm}$ each of this wave type.  These two main shear-induced processes of linear origin (non-modal growth and coupling of waves) ultimately lead to balanced MHD turbulence in shear flows. 

In conclusion, the balancing of MHD turbulence under sufficiently strong, super-Alfv\'enic shear of the flow velocity, revealed in this study, is consistent with recent solar wind observations \cite{Soljento-2023} and can provide an interpretation of these observations within the framework of shear-driven MHD turbulence dynamics.  Of course, in reality, solar wind dynamics involve other complex factors, such as thermodynamics, compressibility and geometric effects, which have not been considered here. However, as noted above, shear often remains one of the main drivers of MHD turbulence. The model of an unbounded flow with constant shear adopted in this work is obviously an idealization, but it does allow us to capture the essence of the dynamics of energy-containing processes driven by shear in more complex flow configurations relevant to the solar wind. Therefore, it is expected that the main results obtained here will also be applicable to more complex shear profiles. In fact, this simplified, yet basic constant shear flow model,  including compressibility (which is ignored here), has previously been used in the solar wind context to study the linear non-modal evolution and shear-induced mutual transformations of Alfv\'en, fast and slow magnetosonic waves \cite{Poedts1998, Gogoberidze2007,Hollweg2013}.  Extending the present study to more realistic solar wind flow models will allow us to clarify how geometry, compressibility, thermodynamics and kinetic effects, such as pressure anisotropy, act together with shear to influence the balance/imbalance properties of MHD turbulence in the solar wind.

\acknowledgments

This work is supported by the Deutsche Forschungsgemeinschaft (DFG) (Grant
No. MA10950/1-1) and Shota Rustaveli National Science
Foundation of Georgia (SRNSFG) (Grant No.  FR-23-1277).
We thank the anonymous Referees for useful comments that improved the presentation of our work.

\bibliography{references}

\end{document}